# Origin of the significant impact of Ta on the creep resistance of FeCrNi alloys

D.Magne[1], X.Sauvage[1*], M.Couvrat[2]

damien.magne@univ-rouen.fr, xavier.sauvage@univ-rouen.fr, mathieu.couvrat@manoir.eu.com

*Corresponding author

[1]Normandie Univ, UNIROUEN, INSA Rouen, CNRS, Groupe de Physique des Matériaux, 76000 Rouen, France
[2] Manoir Industries, 12 Rue des Ardennes - BP 8401 – Pitres, 27108 Val de Reuil, France

**Abstract**

Heat resistant FeCrNi alloys are widely used in the petrochemical industry because they exhibit a unique combination of creep and oxidation resistance at temperatures exceeding 900°C. Their creep properties are often optimized by micro-additions of carbide forming elements. In the present work, the influence of Ta micro-additions has been experimentally investigated both on as-cast and aged microstructures to understand the origin of the significant impact of this element on the creep resistance. Calculations with thermocal software were also carried out to support experimental data. It is shown that a small addition of Ta is beneficial as it increases the volume fraction of stable MC carbides. We demonstrate also that additions of Ta may have a dramatic effect on the thermal stability of microstructures. This is attributed to a smaller equilibrium volume fraction of $M_{23}C_6$ and more pronounced heterogeneous precipitation at MC/matrix interfaces. The influence on the creep properties in then discussed.

## 1. Introduction

Heat resistant FeCrNi steels for applications in the petrochemical industries such as reforming or cracking furnaces have been developed years ago to combine creep and corrosion resistance at temperatures exceeding 900°C [1-2]. Such austenitic steels are iron-based alloys with a high chromium content (typically in a range of 25 to 35wt.%) to resist against corrosion and also a high nickel level (typically in a range of 35 to 45wt.%) to stabilize the austenitic structure. Besides, they contain a significant level of carbon (typically about 0.5wt.%) that creates a relatively large volume fraction of carbides (typically between 4 and 8%vol.) necessary to sustain the stress under creep conditions. Depending on other alloying elements, three types of carbides are usually created during solidification or aging in service conditions: $M_7C_3$, $M_{23}C_6$ and MC. The $M_7C_3$ carbides, where *M* stands mostly for Cr, are stable at high temperature, above application temperatures, and are typically observed in as-cast structures as eutectic carbides [3] or after carburization in a carbon rich environment [4]. During aging in service conditions, they typically transform into the more stable $M_{23}C_6$, where *M* stands also mostly for Cr [5]. Historically, a great improvement of austenitic steels creep properties was achieved through the introduction of Nb in their composition, leading to a significant volume fraction of MC carbides where *M* stands mostly for Nb [3, 6-8]. These MC carbides are stable over a large temperature range and they mostly nucleate during the solidification, thus they are part of typical as-cast structures. Other micro-alloying elements are nowadays also used to optimize the creep properties of commercial heat resistant austenitic steels such as Ti which is known to promote the formation of MC carbides also [9-11]. More recently, it has been proposed to go further using Ta as alloying element [12-15]. An improvement of the creep strength up to 15% [12] is achieved even with a very small amount of Ta. It is important to note first that the underlying mechanisms have not been clarified yet, and second that similarly to Ti micro-additions [9] there is an optimum concentration over which the creep strength or the time to failure in creep test conditions decreases. Thus, the aim of this work was to elucidate the origin of the significant impact of Ta on the creep resistance of such FeCrNi alloys. For this purpose, a series of alloys with a full range of Ta concentration has been investigated. A special emphasis has been given on microstructural evolutions during long time aging and on carbides since they are known to play a key role 0n the creep behavior. Calphad simulations were also used to support experimental data.

## 2. Experimental details

FeCrNi alloys with various Ta content of the present study have been produced by centrifugal casting by Manoir Industries. Their chemical composition is given in table 1. Materials were characterized both in the as-cast state and after aging in atmospheric condition at 1000°C during 100h.

Microstructural evolutions were investigated thanks to Scanning Electron Microscopy (SEM) with a ZEISS LEO 1530 XB operating at 15kV and a back scattered electron (BSE) detector. Samples were cut with a precision saw directly from cast tubes and then mechanically polished with ¼ µm diamond paste prior to SEM observations. For aged materials, an additional chemical etching was performed with an oxalic acid solution to exalt the contrast between primary and fine scale secondary carbides. Carbide volume fractions were estimated from surface fractions calculated by image processing with ImageJ software on images recorded with a magnification of 1000x. One should note however that this procedure has not been applied on images of aged materials because of the high density of fine scaled secondary carbides that could not always be captured with a good accuracy.

To support the experimental data, some simulations were carried out by CALPHAD calculation with the Thermocalc 2017-a software using the TCFE8 (v8.1) data base. Only calculations at the thermodynamic equilibrium could be done, which is in principle not the case of as-cast structures. Anyway, to obtain estimates of expected phases and their composition in as-cast conditions, calculations were done at a temperature of T=1200°C which is about 100°C below the solidus line.

| Element | C | Mn | Si | Ni | Cr | Mo | Nb | W | Ti | Fe | Ta |
|---|---|---|---|---|---|---|---|---|---|---|---|
| wt.% | 0.45-0.5 | 0.5-1.5 | 1.0-2.0 | 33.0-36.0 | 24.0-27.0 | 0.0-0.5 | 0.5-1.0 | 0.15-0.50 | 0.05-0.15 | Bal. | 0-0.2 |

Table 1: Chemical composition (wt.%) of the alloys studied in this present work.

# 3. Results and discussion

## 3.1 Influence of Ta on as-cast microstructures

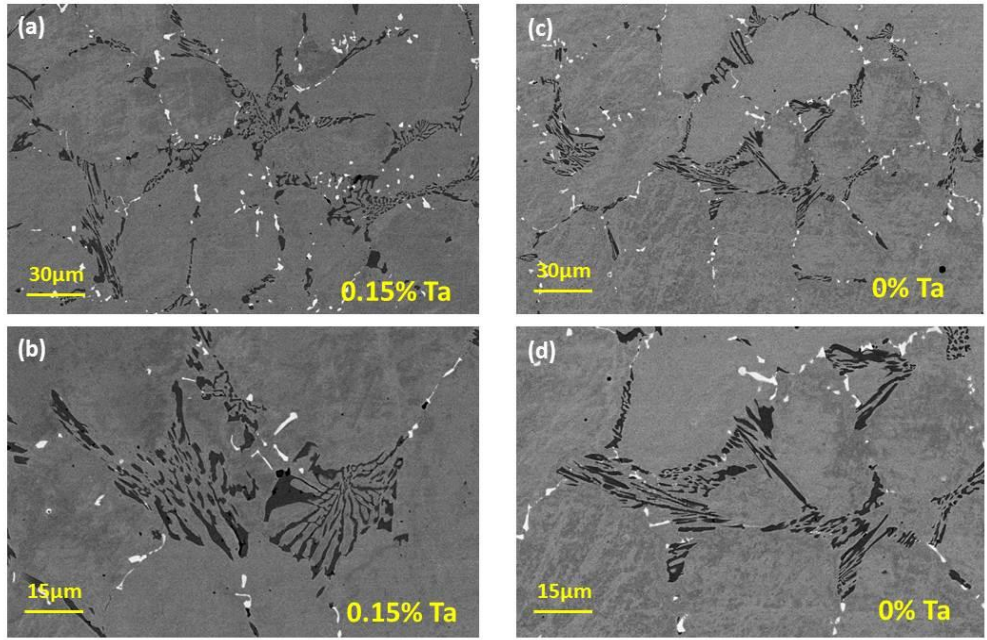

*Figure 1*

*Figure 1: SEM images (BSE detector) of as-cast structures at two different magnifications. FeCrNi alloy with 0.15 wt.% Ta (a) and b)) , FeCrNi alloy with 0 wt.% Ta ((c) and d)).*

A large number of alloys with a full range of Ta has been investigated, however to highlight the influence of Ta on microstructures only SEM images recorded for two alloys, one without Ta and the other with 0.15wt.%Ta, are shown here. As-cast microstructures (Fig.1) clearly exhibit relatively large equiaxed areas with a uniform contrast corresponding to dendrite arms formed during the solidification. They are surrounded by eutectic areas that exhibit three different contrasts: bright particles are MC carbides, dark particles are $M_7C_3$ carbides and grey zones are the austenitic matrix. Such as-cast structure is absolutely typical of this kind of FeCrNi alloys [2, 9, 11], and interestingly micro-addition of 0.15wt.% Ta does not lead to any significant visible difference. Nevertheless, since this element is expected to promote the formation of MC carbides, the volume fraction of these carbides in the as-cast state has been precisely quantified by image processing for all alloys. The plot of Fig. 2a clearly reveals a systematic increase of the volume fraction of MC carbides with the Ta content with a linear dependence (from about 0.8 to 1.2 vol. % for 0 to 0.18wt.% Ta respectively).

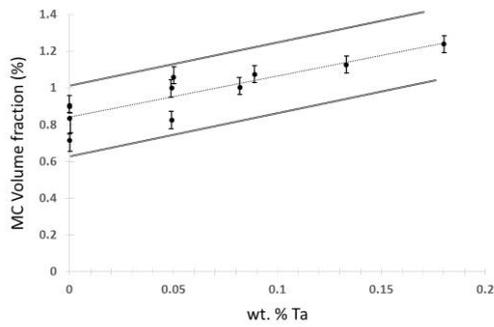 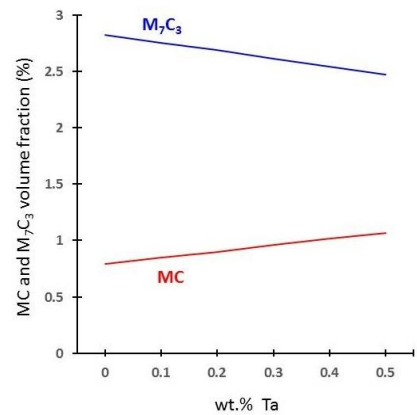

*Figure 2a: Experimental volume fraction of MC phase in the as-cast state as a function of Ta concentration (wt.%). Data extracted from SEM image processing.*

*Figure 2b: Volume fractions (vol. %) of $M_7C_3$ and MC carbides as predicted by thermocalc at 1200°C for the FeCrNi alloys of the present study and plotted as a function of the Ta concentration (wt.%)*

The predictions given by thermocalc at 1200°C (Fig. 2b) are relatively in good agreement with the experimental data for the alloy without Ta (MC volume fraction of about 0.8%). It indicates that the temperature that has been arbitrary chosen to account for the as-cast structure (1200°C) provides some realistic estimates. The MC volume fraction increases linearly with the Ta content as observed experimentally (Fig. 2a) but the slope is smoother, rising only up to about 0.9% for 0.18wt.%Ta. Obviously, the increase in MC volume fraction is compensated by a decrease of $M_7C_3$ (Fig. 2b) resulting from the limited availability of carbon in the alloy. Besides, it is important to note that thermocalc calculations did not reveal any significant variation of the composition of carbides when the Ta amount is changed (data not shown here). Thus, in summary, even if SEM images (Fig. 1) do not exhibit any significantly different microstructures, one should keep in mind that micro-additions of Ta slightly affect the volume fraction of primary carbides (MC and $M_7C_3$).

*3.2 Influence of Ta on microstructures aged at high temperature*

After 100h at 1000°C, the original as-cast microstructures of the FeCrNi alloys without and with 0.15wt.% Ta have deeply changed (Fig. 3). In both alloys, lamellar eutectic zones have disappeared and during aging, primary carbides have evolved toward more equiaxed shapes. It is also very interesting to note that the FeCrNi alloy that contains 0.15wt.% Ta exhibit significantly less (if any) secondary carbides. Secondary $M_{23}C_6$ carbides typically results from the transformation of the unstable large as-cast $M_7C_3$ carbides into $M_{23}C_6$ leading to some carbon release in the matrix and homogeneous precipitation of fine scaled additional $M_{23}C_6$ in the matrix during aging [2, 6, 15]. In the present case, it is clear that Ta micro-additions deeply affect this process; leading to a very different distribution and morphology of $M_{23}C_6$ carbides after 100h at 1000°C (they are more globular in the alloy without Ta, see Fig. 3b and 3d).

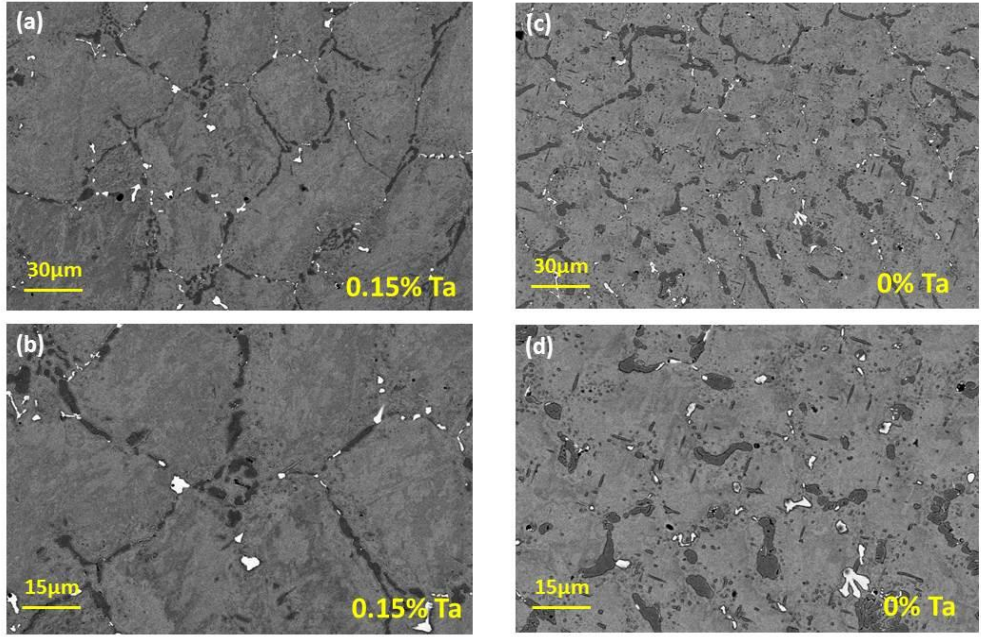

*Figure 3: SEM images (BSE detector) of annealed structures (100h at 1000°C-) at two different magnifications. FeCrNi alloy with 0.15 wt.% Ta (a) and b)) , FeCrNi alloy with 0 wt.% Ta ((c) and d)).*

The equilibrium volume fractions of carbides at 1000°C estimated by thermocalc are plotted as a function of the Ta content on the Fig.4. As expected, only MC and $M_{23}C_6$ carbides are stable. The volume fraction of MC is close to that at 1200°C, and it also increases linearly as a function of the Ta content. One should note however that the volume fraction of $M_{23}C_6$ is significantly larger than the fraction of initial $M_7C_3$ (Fig. 2b), which is attributed to the much larger M/C ratio in $M_{23}C_6$. Besides, the equilibrium volume fraction of $M_{23}C_6$ decreases when the Ta content is increased. Thus, in summary, Ta micro-addition affects the equilibrium volume fraction of carbides at a temperature near service conditions, and it strongly affects microstructural evolutions upon aging, especially secondary precipitation and carbide coarsening and globalization. Then, considering the critical role played by these carbides on creep resistance [1, 6-9], these data make it possible to understand the impact of Ta on creep properties. A small amount is beneficial as it leads to an increase of the volume fraction of MC carbides that are more stable at high temperature and less prone to coarsening [3, 6-8]. However, if the Ta concentration is too high, it affects significantly the secondary precipitation in the austenitic matrix and the creep resistance decreases [2, 7]. An interesting question is however left open: why and how does the small Ta content could affect the microstructure evolution during aging at high temperature?

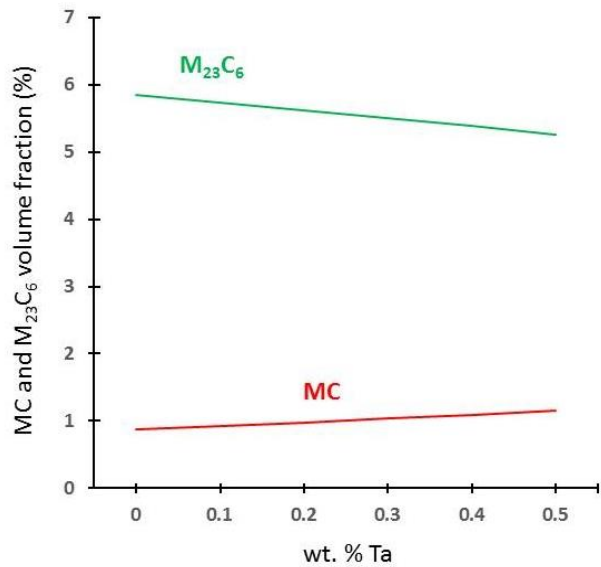

*Figure 4: Volume fractions (vol. %) of $M_{23}C_6$ and MC carbides as predicted by thermocalc at 1000°C for the FeCrNi alloys of the present study and plotted as a function of the Ta concentration (wt.%)*

### 3.3 Understanding the role of Ta on microstructure evolutions at high temperature

As mentioned before, primary $M_7C_3$ carbides are unstable at service temperature (Fig. 4) and transform into $M_{23}C_6$ carbides at 1000°C. Since the M/C ratio of these two carbides is significantly different (2.33 vs 3.83 for $M_7C_3$ and $M_{23}C_6$ respectively), this transformation leads to a significant release of C atoms in the matrix promoting the homogeneous nucleation of a fine dispersion of secondary $M_{23}C_6$ carbides [2]. These carbides easily nucleate and grow since they exhibit a cube on cube orientation relationship with the fcc matrix with only a small lattice mismatch [5]. However, as shown on the SEM image of Fig. 5a and schematically represented on Fig. 5b, some heterogeneous nucleation at MC/matrix interfaces also occurs. It is important to note that it has been observed for any alloy of the present study, thus it is not a chemical effect due to Ta micro-additions. Primary MC carbides have no specific orientation relationship with the matrix, so these interfaces are incoherent and are ideal nucleation sites for $M_{23}C_6$ carbides. It is also worth mentioning that primary MC carbides are usually located in eutectic inter-dendritic zones, i.e. near primary $M_7C_3$ carbides that act as carbon reservoir. Then, since Ta micro-additions lead to an increase of primary MC carbides (Fig. 2b), then the number of heterogeneous nucleation sites increases with Ta concentration. Simultaneously, it also decreases the $M_{23}C_6$ equilibrium volume fraction (Fig.4) and these two effects significantly decrease the volume fraction of homogeneously precipitated $M_{23}C_6$. Moreover, the stability against coarsening of heterogeneously nucleated carbides is probably higher. Indeed, dissolving some of these carbides (which is necessary to coarsen others) leads to the re-formation of MC/matrix interfaces which is not the case for carbides dissolved inside the matrix.

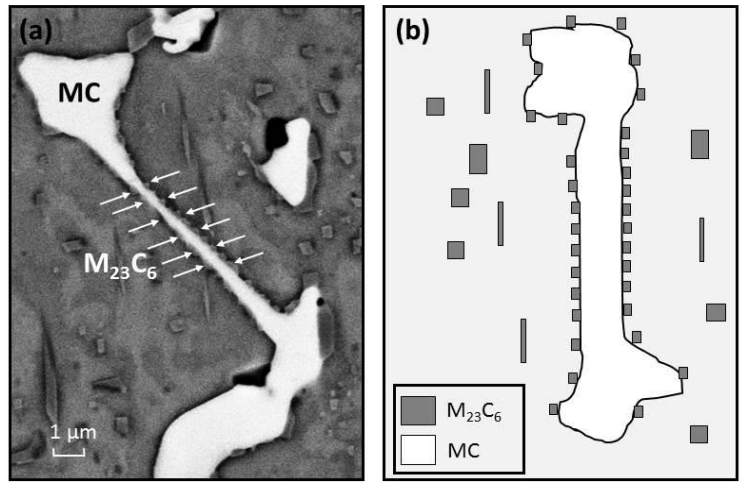

*Figure 5: (a) SEM image (BSE) of the FeCrNi alloy with 0.15wt.% Ta aged at 1000°C during 100h. Heterogeneous nucleation of secondary $M_{23}C_6$ carbides at the MC/matrix interface is arrowed. (b) Schematic representation of the secondary precipitation that occurs both in the matrix and on the surface of MC carbides.*

Thus, adding Ta has a small effect on as-cast microstructures, slightly affecting volume fractions of primary carbides, however it has a series of consequences on microstructure evolution at high temperature as schematically represented on Fig. 6: it creates at first more heterogeneous nucleation sites for $M_{23}C_6$, ii) it decreases the equilibrium volume fraction of $M_{23}C_6$ and thus the fraction of homogeneously nucleated secondary carbides, iii) heterogeneously precipitated $M_{23}C_6$ carbides being more stable, Ta also indirectly reduces coarsening and globalization. This is of course more pronounced for higher Ta level and it explains why an excessive concentration of Ta affect the microstructure evolution at high temperature and thus the creep resistance of heat resistant austenitic FeCrNi alloys.

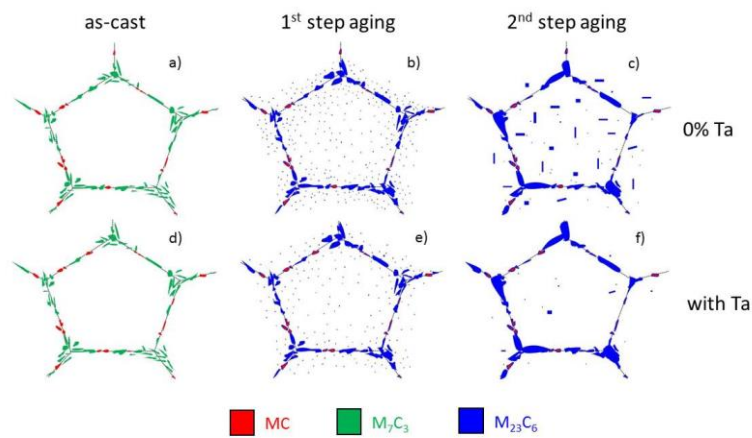

*Figure 6: Schematic representation of the microstructural evolution during aging at high temperature of FeCrNi alloys without Ta (a, b, c) and with micro-addition of Ta (d, e, f). The as cast structure (a, d) gives rise during a first step to secondary precipitation (b, e) and then significant coarsening of $M_{23}C_6$ occurs (c, f).*

## 4. Conclusions

i) Ta micro-additions lead to an increase of the MC carbide volume fraction and a decrease of the $M_7C_3$ carbide volume fraction in the as-cast state.

ii) Although little differences are visible in the as-cast states, alloys without and with 0.15wt.% exhibit very different microstructures after aging at 1000°C during 100h.

iii) The homogeneous secondary precipitation of $M_{23}C_6$ is reduced when Ta is added, resulting from a reduced flow of carbon resulting from the transformation of primary $M_7C_3$ and to heterogeneous precipitation at MC/matrix interfaces.

iv) Heterogeneously nucleated $M_{23}C_6$ carbides being thermally more stable, large micro-additions of Ta lead to heterogeneous distribution of $M_{23}C_6$ carbides after aging at 1000°C during 100h.

v) There is an optimum Ta concentration for creep properties. It results from the balance between the positive impact of increased volume fraction thermally stable MC carbides and the detrimental effect on the volume fraction and the thermal stability of homogeneously nucleated $M_{23}C_6$.

**Acknowledgements**

This work has been funded by the Agence National de la Recherche (ANR), project IPERS, grant number LAB COM – 15 LCV4 0003.